\documentclass{aa}
\usepackage{graphics} 
\begin{document}

\thesaurus{07.(02.08.1;02.20.1;08.09.3)}

\title{The small-P\'eclet-number approximation in stellar radiative zones}
\author{F. Ligni\`eres}
\institute{D\'epartement de Recherche Spatiale et Unit\'e de Recherche
associ\'ee au CNRS 264, Observatoire de Paris-Meudon, F-92195
Meudon Cedex, France (internet:Francois.Lignieres@obspm.fr)}
\date{Received 15 September 1998 / Accepted 21 May 1999}
\maketitle
\begin{abstract}
We present an asymptotic form of the Boussinesq equations in the limit of
small P\'eclet numbers i.e. when the time scale of motions is much larger than 
the time scale of thermal diffusion. We find that, in this limit, 
the effects of 
thermal diffusion and stable stratification combine in a single physical
process. This process is an anisotropic dissipation (not effective for 
horizontal motions) which acts primarily on large scale motions. The 
small-P\'eclet-number approximation presents also the great practical interest to avoid
the numerical difficulty induced by the huge 
separation between the diffusive and dynamical time scales. 
The relevance of this approximation to study the flow dynamics within the 
stellar radiative zones is considered.

\end{abstract}

\section{Introduction}\label{sec:level1}

The understanding of the flow dynamics within stellar radiative zones 
constitutes a major challenge for the current theory of stellar evolution. 
These motions transport chemical elements and it turns out that their 
contribution might reconcile the existing models of stellar structure with the 
observations of the surface abundances (Pinsonneault \cite{pin}). 
By transporting angular 
momentum, such flows also play an important role in the evolution of star's 
rotation. In particular, they could explain the nearly solid body rotation of 
the solar radiative zone which has been revealed by 
helioseismology (Gough et al. \cite{gough}).

We consider here the effects of the very high thermal
diffusivity of stellar interiors on the dynamics of these motions. 
In most cases, radiation dominates the thermal 
exchanges within radiative zones. This heat transport is so 
efficient that the thermal diffusivities associated with the radiative flux 
are larger by several orders of magnitude than the thermal diffusivities 
encountered in colder media like planetary atmospheres. For example, the 
thermal diffusivity varies between $10^5$ and $10^7 \;{\rm cm}^2 
{\rm s}^{-1}$ inside the sun whereas it is equal to $0.18\;{\rm cm}^2 
{\rm s}^{-1}$ in the standard conditions of the terrestrial atmosphere.

This property of the stellar fluid is expected to strongly affect the flow 
dynamics especially inside the stably 
stratified radiative zone where the time scale of thermal diffusion appears 
to be shorter than the dynamical time scale characterizing radial motions.
Helioseismology data show that the thermal structure of this 
region is very close to the one predicted by hydrostatic models, indicating 
that existing fluid motions are not fast enough to 
modify significantly the thermal structure built up by the radiative flux
(Canuto \& Christensen-Dalsgaard \cite{can}). 

Qualitatively, the damping of temperature 
fluctuations by thermal diffusion is expected to have two main effects on the 
dynamics. The first one is to reduce the amplitude of the buoyancy force. This 
restoring force acts on fluid parcels displaced from their equilibrium level 
and is proportional to the density difference between the parcel and its 
environment. Since density fluctuations are proportional to temperature 
fluctuations for incompressible motions, fast thermal exchanges reduce the 
force amplitude. An important consequence of this effect is to favour the onset
of shear layer instabilities in stably stratified layers (Dudis \cite{dudis}, 
Zahn \cite{zan}). 

The second main effect of the thermal diffusion
is to increase the dissipation of kinetic energy. Any vertical motion in
a quiescent atmosphere induces a work of the buoyancy force so that 
a fraction of the injected kinetic energy is necessarily transformed
into potential energy. If the fluid parcels could "fall" adiabatically 
towards their equilibrium position, all the stored potential energy could
return back to kinetic energy. However, the damping of temperature fluctuations
provokes an irreversible loss of kinetic energy. A simple example of this
process is the damping of gravity waves.

Both effects of the thermal diffusivity are thus opposed. While a decrease of
the buoyancy force amplitude reduces the associated work and thus the amount
of kinetic energy extracted, the second effect increases the fraction of the 
kinetic energy which is irreversibly lost. Then, for a given mechanical 
forcing, a relevant question is whether
a larger thermal diffusivity reduces or enhances the kinetic energy of the 
flow.

We are lacking quantitative results especially in non-linear regimes to answer 
such a basic question and more generally to understand the effect of thermal 
diffusivity in a stellar context. This situation is partly due to the 
difficulty to reproduce flows with realistic Prandtl numbers either
in laboratory experiments or by numerical simulations. The Prandtl number 
$P_r =\nu /\kappa$ which compares the kinematic viscosity $\nu$ and the thermal 
diffusivity $\kappa$, varies between $10^{-6}$ and $10^{-9}$ within the sun 
whereas it is equal to $0.7$ in the air. Although some fluid like metal 
liquid may have small Prandtl numbers in laboratory conditions ($P_r =0.025$ for 
mercury, see for example Cioni et al. \cite{som})),
these values remains far from the stellar case.
The severe 
numerical limitation is explained by the huge separation
between the time scales of viscous dissipation and thermal diffusion. The 
computation of both processes over a few dynamical times would require a 
prohibitive amount of computer time.

In this paper, we investigate the limit where the time scale characterizing the 
thermal exchanges is much shorter than the time scale of the motions (the ratio
between both time scales defines the P\'eclet number). In Sect. 2, 
an asymptotic form of 
the governing equations is derived in the context of the
Boussinesq approximation. Evidences that these asymptotic equations
actually approximate the Boussinesq equations for small P\'eclet numbers
are presented in Sect. 3. Then, in Sect. 4, the elementary properties
of the small-P\'eclet-number equations are described, 
emphasizing their theoretical and practical interests.
Finally, the relevance of
this approximation in a stellar context is commented in Sect. 5.

\section{Derivation of the small-P\'eclet-number approximation}
\label{sec:level2}

We restrict ourselves to a fluid layer embedded in an uniform vertical gravity field
and bounded by two horizontal plates. A mechanical forcing is assumed to drive 
motions which can be described by the Boussinesq approximation. We do not need to 
specify the forcing for the moment, we only assume that it introduces a velocity scale $U_*$.
The temperature is fixed on both plates so that a linear diffusive profile denoted $T^{i}(z)$ is
established initially. The dynamical effect of the stable stratification is measured by 
the Brunt-V\"{a}is\"{a}l\"{a}  frequency, 
$N_* = \left(\beta g {\Delta T}_*/ L_* \right)^{1/2}$, where $g$ 
denotes the gravity acceleration, $\beta$ is the thermal expansion coefficient,
${\Delta T}_*$ the temperature difference between the upper and lower plates 
and $L_*$ the distance separating the plates. 

In the context of the Boussinesq approximation, the governing non-dimensional 
equations read:

\begin{equation} \label{eq:vel1}
\frac{\partial {\bf u}}{\partial t} + {\bf u}\cdot \nabla {\bf u} =
- \nabla p + R_i \theta {\bf e}_z + \frac 1{R_ e} \nabla ^2{\bf u},
\end{equation}

\begin{equation} \label{eq:temp1}
\frac{\partial \theta}{\partial t} + {\bf u}\cdot \nabla \theta + w =
\frac1{P_ e}\nabla ^2 \theta,
\end{equation}

\begin{equation} \label{eq:div1}
\nabla \cdot {\bf u}=0,
\end{equation}

\noindent
where, ${\bf u} = u {\bf e}_{x} + v {\bf e}_{y} + w {\bf e}_{z}$ is the 
velocity vector, $p$ the pressure and $\theta(x,y,z) = 
T(x,y,z) - T^{i}(z)$ the temperature deviation from the initial
temperature profile. The $z$ axis refers to the vertical direction, while the $x$ and $y$ 
axis refer to the horizontal ones. 
In the heat equation, the third term of the left
hand side 
corresponds to the vertical advection of temperature against the mean 
temperature gradient $d T^{i}(z)/dz$. This gradient is equal to unity in the 
dimensionless unit.
To non-dimensionalize the equations we used the velocity scale $U_*$, the
length scale $L_*$, the dynamical time scale $t_{\rm D} = L_*/U_*$, the pressure scale
$\varrho_0 U_*^2$ and the temperature
variation ${\Delta T}_*$. 

The system is then governed by the Richardson number, $R_ i$, the P\'eclet number, 
$P_e $, and the Reynolds number, $R_e$, respectively defined as 
\[ R_i = \left(\frac{N_* L_*}{U_*}\right)^{2}, \;\;\; 
P_e = \frac{U_* L_*}{\kappa} ,\;\;\; R_e = \frac{U_* L_*}{\nu}.\]
\noindent
The Richardson number is the square of the ratio between the dynamical time
scale $t_{\rm D}$ and the buoyancy time scale $t_{\rm B}= 1/N_*$. 
The thermal diffusivity $\kappa$ 
appears in the P\'eclet number which compares the thermal diffusion time scale, 
$t_{\kappa} = L_*^2/ \kappa$ with the dynamical time scale. The Reynolds number 
is the ratio between the viscous time scale $L_*^2/ \nu$ and the dynamical time
scale.

In the limit of small P\'eclet number, we assume that the solutions 
${\bf u}$ and $\theta$ of the Boussinesq equations behave like
Taylor series:
\begin{equation} \label{eq:devv}
{\bf u} = {\bf u_0} + P_e {\bf u_1} + P_e^2 {\bf u_2} + ...
\end{equation}
\begin{equation} \label{eq:devt}
\theta = \theta_0 + P_e \theta_1 + P_e^2 \theta_2 + ... .
\end{equation}
\noindent
Note that in the context of the Boussinesq equation, the pressure is an intermediate
variable determined by the incompressibility condition (\ref{eq:div1}).
By inserting these asymptotic 
expansions in the heat equation, we find at the zero order in $P_e $:

\begin{equation} \label{eq:temp0_N}
\nabla ^2 \theta_0 = 0.
\end{equation}
\noindent
Since the temperature remains fixed to its initial value on both bounding plates,
temperature deviations vanish on both plates. Then,
Eq. (\ref{eq:temp0_N}) implies 

\begin{equation} \label{eq:temp0}
\theta_0 = 0.
\end{equation}
\noindent
Thus, at the lowest order in $P_e$, the Boussinesq equations reduce
to the Navier-Stokes equation:

\begin{equation} \label{eq:vel0}
\frac{\partial {\bf u_0}}{\partial t} + {\bf u_0}\cdot \nabla {\bf u_0} =
- \nabla p_0 + \frac 1{R_e} \nabla ^2{\bf u_0},
\end{equation}

\noindent
together with the incompressibility condition,

\begin{equation} \label{eq:div0}
\nabla \cdot {\bf u_0}=0.
\end{equation}

\noindent
At this order, the dynamical and thermal equations are decoupled.
The coupling is recovered at the first order in $P_e$:

\begin{equation} \label{eq:vel2}
\frac{\partial {\bf u_1}}{\partial t} + {\bf u_0}\cdot \nabla {\bf u_1} 
+ {\bf u_1}\cdot \nabla {\bf u_0} =
- \nabla p_1 + R_i \theta_1 {\bf e}_z + \frac 1{R_e} \nabla ^2{\bf u_1},
\end{equation}

\begin{equation} \label{eq:temp2}
w_0 = \nabla ^2 \theta_1,
\end{equation}

\begin{equation} \label{eq:div2}
\nabla \cdot {\bf u_1}=0.
\end{equation}

\noindent
Solutions ${\bf \hat{u} }= {\bf u_0} + P_e {\bf u_1}, \hat{\theta} = \theta_0 + 
P_e \theta_1$ valid 
up to the first order in $P_e$ must satisfy the above system of equations
(\ref{eq:temp0}), (\ref{eq:vel0}), (\ref{eq:div0}), (\ref{eq:vel2}), (\ref{eq:temp2}), (\ref{eq:div2}).

We note that the Lagrangian derivative of temperature deviations does not
appear in the heat equation of this system.
Thus, at the first order in $P_e$, one would have found the same system of equations for 
${\bf u_0}$, ${\bf u_1}$, $\theta_0$, $\theta_1$ if the Taylor series had been introduced 
in the following equations:

\begin{equation} \label{eq:vela0}
\frac{\partial {\bf u}}{\partial t} + {\bf u}\cdot \nabla {\bf u} =
- \nabla p + R_i \theta {\bf e}_z + \frac 1{R_e} \nabla ^2{\bf u},
\end{equation}

\begin{equation} \label{eq:tempa0}
P_e w = \nabla ^2 \theta,
\end{equation}

\begin{equation}  \label{eq:diva0}
\nabla \cdot {\bf u}=0.
\end{equation}

\noindent
Therefore, if ${\bf u}$ and $\theta$ actually behave as Taylor series
for small P\'eclet numbers, the solution of the above equations is 
identical to the solution of the Boussinesq equations up to the first order in $P_e$.

The unique difference with the Boussinesq equations comes from the heat 
equation.
Physically, the process leading to the balance 
$P_e w = \nabla ^2 \theta$ can be described as follows:
For large values of the thermal diffusivity, 
the temperature fluctuations are expected 
to be small and the mean temperature stratification to remain unchanged
by the mechanical heat flux. 
However, vertical motions advecting fluid parcels against the 
mean temperature gradient always produce temperature deviations and,
unlike the non-linear advection term ${\bf u}. \nabla \theta$,
this generation process does not depend on the amplitude of the temperature
deviations.
As fluid parcels go up (or down) in a mean temperature gradient,
the amplitude of the temperature deviations tends to increase
continuously. In the mean time, thermal diffusion tends to reduce these
temperature deviations. Inspection of the heat Eq. ({\ref{eq:temp1})
shows that this diffusive process can lead to a stationary solution, namely
$P_e w = \nabla^2 \theta$. Clearly, if the time scale of the vertical motions is very slow 
compared to the diffusive time scale, one expects that this stationary solution is 
practically instantaneously reached. Again, it describes a balance between thermal
diffusion and vertical advection against the mean temperature stratification.

In the remainder of this paper, we will refer to the set of equations (\ref{eq:vela0}),
(\ref{eq:tempa0}), (\ref{eq:diva0}), as
the small-P\'eclet-number equations or as the small-P\'eclet-number approximation.
However, formal mathematical proof that ${\bf u}$ and $\theta$ actually behave
as Taylor series does not exist in the general case. 
Then, to prove that the small-P\'eclet-number equations
actually approximate the Boussinesq equations in the
limit of small P\'eclet number, specific cases have to be considered.
In the next section, we shall present two types of linear flows where
the validity of the small-P\'eclet-number approximation can be proved.
Some evidences will also be given for a non-linear flow.
The theoretical and practical interest of the small-P\'eclet-number equations
will be emphasized in Sect. 4.

\section{Validity of the small-P\'eclet-number approximation}

The first example we consider is that of small amplitude perturbations in a linearly
stably stratified atmosphere. 
The perturbations are resolved into modes proportional to 
${\rm exp}(-\sigma t) {\rm exp}[i(k_x x + k_y y + k_z z)]$ where $\sigma$ is a complex number
and $k_x$, $k_y$, $k_z$, represent the horizontal and vertical wave numbers of the 
perturbation. In the following, 
the dispersion relation obtained using the Boussinesq 
equations is compared to that derived from the 
small-P\'eclet-number equations.

The calculation is conducted for two dimensional disturbances ($k_y =0$), but the
three-dimensional case can be readily recovered replacing $k_x^2$ by 
$k_x^2 + k_y^2$ in the following expressions. 
To simplify the presentation we also limit ourselves to the inviscid case. It has
been verified that our conclusions are not affected
by taking into account the viscosity.

Using the Boussinesq equations, the dispersion relation is:
\begin{equation} \label{eq:disp1}
\sigma^2 - \sigma_T \sigma + \sigma_B^2 = 0
\end{equation}
\noindent
whereas the dispersion relation reduces to
\begin{equation}
\sigma = \frac{\sigma_B^2}{\sigma_T}
\end{equation}
\noindent
in the context of the small-P\'eclet-number equations.
In these expressions, 
\[
\sigma_T  = \frac{k_x^2 + k_z^2}{P_e}
\]
\noindent
is the damping rate associated with a pure thermal diffusion, and
\[
\sigma_B = \sqrt{R_i} \frac{k_x}{\sqrt{k_x^2 + k_z^2}}
\]
\noindent
is the frequency of gravity waves in absence of diffusive processes.

We observe that, in the context of the small-P\'eclet-number approximation,
all disturbances are damped with a rate equal to 
$\sigma_B^2 / \sigma_T$.
On the contrary, the dispersion relation of the Boussinesq equations shows 
different types of solutions. These solutions are now analyzed 
for increasing values of the thermal diffusivity.

At small thermal diffusivity, solutions of the dispersion relation
correspond to gravity waves damped by thermal diffusion.
The two roots of Eq. (\ref{eq:disp1}) correspond to  two gravity 
waves propagating in opposed direction.
Increasing the thermal diffusion reduces the wave frequency until the roots 
of (\ref{eq:disp1})  become purely real and the associated modes 
damped without propagating. This occurs when
\[
P_e < \frac{{\left(k_x^2 + k_z^2 \right)}^{3/2}}{2 \sqrt{R_i} k_x}.
\]
\noindent
It is important to note that, as long as $k_z$ is not equal to zero, there always exists
a P\'eclet number such that this expression is verified for all $k_x$ and $k_z$.
If this was not the case, one could have gravity waves 
whatever the value of $P_e$. Then, 
the small-P\'eclet-number approximation would not be valid
since gravity waves are absent in this approximation.

In deriving the small-P\'eclet-number equations, we restricted ourselves
to a fluid layer bounded vertically. Vertical wave numbers have 
therefore a lower limit, $k_z^{min}$, so that
all modes are damped without propagation if 
$P_e < 3 \sqrt{3} k_z^{min} / 4 \sqrt{R_i}$.

Then, the two distinct roots 
of the dispersion relation correspond to two damping modes. 
By further increasing the diffusivity, the
damping rates take increasingly different values 
and the associated modes 
correspond to two different types of motions.

In the limit of small P\'eclet numbers, the damping rate of the first type of mode is:
\[
\sigma = \frac{\sigma_B^2}{\sigma_T} =  
R_i P_e \frac{k_x^2}{{\left( k_x^2 + k_z^2 \right)}^{2}}
\]
\noindent
These are exactly the weakly damped modes 
found in the context of the small-P\'eclet-number approximation. 
Note that, despite the high thermal diffusivity, 
temperature perturbations can be weakly damped,
if they are associated with
vertical motions against the mean temperature gradient.

The damping rate of the second type of mode is:
\[
\sigma = \sigma_T  = \frac{k_x^2 + k_z^2}{P_e}
\]
\noindent
Such modes are not found in the context of the small-P\'eclet-number
equations. Note that this is not surprising since they correspond to 
solutions of the Boussinesq equations which do not behave like Taylor series (see
equations (\ref{eq:devv}) and (\ref{eq:devt})) when the P\'eclet number goes to zero.
These modes undergo a purely diffusive damping 
which can be made
arbitrarily fast as the P\'eclet number vanishes.
Indeed, whatever the values of $k_x$ and $k_z$,
all these modes
are reduced by an arbitrary large factor after a time proportional to
${P_e}/k_z^{min}$.
For this type of motions, the vertical advection term appearing in the 
linearized heat Eq. (\ref{eq:temp1}) is negligible. This shows that, 
in the limit of small P\'eclet number, temperature perturbations which 
are not produced by vertical advection are damped in a very short time.

According to the above discussion, it is always possible to find
a P\'eclet number such that, after an arbitrarily small time, 
the evolution of the infinitesimal perturbations is equally described by 
the Boussinesq equations or by the small-P\'eclet-number equations.

We now consider another example of flow, yet in a linear regime. It concerns
the evolution of small disturbances in a stably stratified shear layer. This
configuration differs from the previous example by the presence of a mean horizontal
flow sheared in the vertical direction. 
In this case, the validity of the linear version of the
small-P\'eclet-number approximation
has already been proved by Dudis' theoretical work (1974).
This author considered specifically 
an hyperbolic-tangent velocity profile
in a stable atmosphere characterized
by a hyperbolic-tangent temperature 
profile and used a normal mode approach 
to study the stability of the flow.

He first determined the 
neutral stability curve, i.e. the curve separating the stable
and unstable regions in the parameter space, for decreasing 
values of the P\'eclet number. Then, 
he showed that for small P\'eclet
numbers these neutral curves could be recovered
using a linear version of the small-P\'eclet-number equations.
The convergence of the Boussinesq equations towards the small-P\'eclet-number 
equations appears fairly rapid in this case since,
already at $P_e =0.2$, the maximum difference between the neutral curves is 
within $3$ percent. 
We recently revisited the work of Dudis by considering a linear 
temperature profile instead of the tangent hyperbolic profile to
characterize the stable stratification (Ligni\`eres 
et al. \cite{moibis}). We confirmed the validity of the  
small-P\'eclet-number equations to describe
the neutral curves. 
In addition, 
we verified its
validity for other types of mode (unstable modes symmetric to the shear
layer mid-plane) as well as in the viscous case.
Note that very rapidly damped modes corresponding to the second type
of mode found in the previous discussion may also exist in this case.
However, they can not affect the stability of the shear layer
since they are very strongly damped.

The third example is a two-dimensional non-linear flow where a shear layer is 
forced at the top of a linearly stratified fluid. 
This flow has been studied 
numerically by Ligni\`eres et al. (\cite{moi})
for large Reynolds numbers ($R_e 
\approx 2000$ where $R_e$ is based on the layer thickness and the velocity 
difference across it). Figure 1a shows the typical vorticity field resulting 
from the destabilization of the shear layer and the concentration of vorticity 
into vortices. The mean shear and the thermal stratification are also 
represented in Figs. 
1b and 1c (here, the means refer to horizontal averages).

The other parameters being held fixed, we reduced the P\'eclet number from 
$1000$ to $1$ (equivalently the Prandtl number $P_r$ has been decreased from 
$0.5$ to $5 \times 10^{-4}$ which already requires some computational effort).
Figures 1d and 1e present horizontal profiles of the vertical velocity and the 
temperature deviation for the two extreme values of the P\'eclet number. When 
this number is equal to $1000$ (Fig. 1d), one recovers a classical property 
of the inflexional shear layer instability in stably stratified medium, namely 
that the phase lag between the vertical velocity and the temperature deviation 
is $\pi/2$. By contrast, we observe that both fields are antiphased when the 
P\'eclet number is equal to unity (Fig. 1e). This striking difference reveals
a change in the predominant terms of the heat equation and we verified that 
this equation is now dominated by a balance between the vertical advection 
against the mean temperature gradient and the thermal diffusion. These 
first results are consistent with the convergence of the Boussinesq equation 
towards the small-P\'eclet-number approximation. A detailed comparison of the results 
of these simulations with those obtained by using the asymptotic equations
will be reported in a forthcoming paper.

\begin{figure*}
\resizebox{\hsize}{!}{\includegraphics{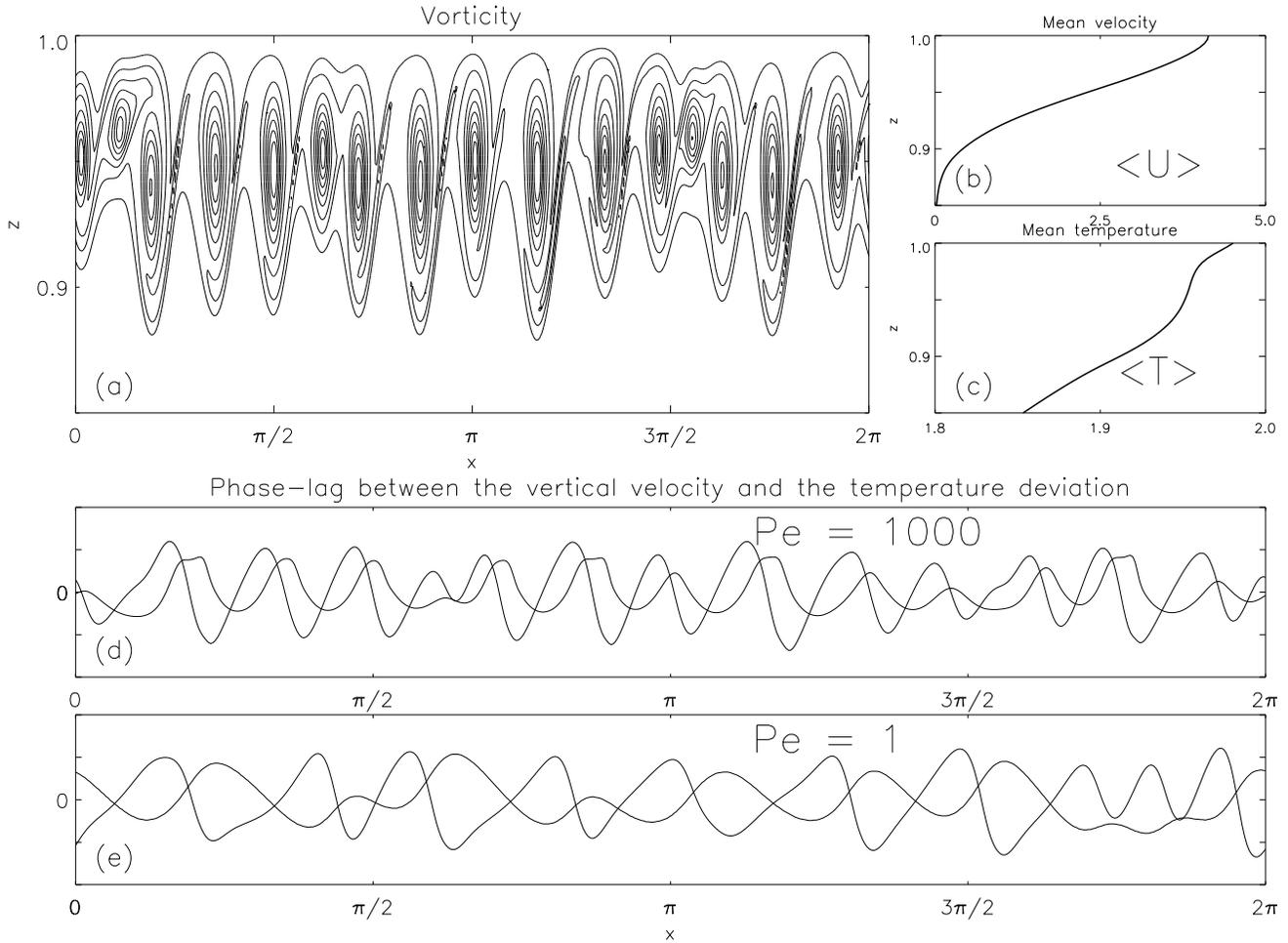}}
\caption{Effect of the P\'eclet number on a non-linear two-dimensional shear
flow. The figures (a), (b), (c), show a typical vorticity field of the flow and
the vertical profiles of the mean horizontal velocity 
and of the mean temperature, respectively. Horizontal variations of  
the vertical velocity and of 
the temperature deviation are displayed for two values of the P\'eclet number,
$P_e = 1000$ (d) and $P_e =1$ (e). Both quantities are antiphased  
in the low
P\'eclet number case, this result pointing towards the validity of the 
small-P\'eclet-number approximation}
\end{figure*}

In this section, the validity of the
small-P\'eclet-number approximation 
has been proved for two types of linear flows.
For the case of infinitesimal perturbations in a linearly stably stratified atmosphere,
we noted that vertical length scales of perturbations have to be
limited to finite values to ensure uniform convergence. Moreover,
the Boussinesq and the small-P\'eclet-number equations 
give the same solution after an arbitrarily small time,
once initial temperature perturbations not associated with
vertical advection against the mean temperature gradient have been damped.
The example of a non-linear flow we considered is also consistent with the
validity of the approximation.

\section{Elementary properties of the small-P\'eclet-number approximation}

The elementary 
properties of the small-P\'eclet-number equations are analyzed in this section.
From a practical point of view, the main interest of these equations is that their numerical
integration does not require the computation of the very rapid temporal variation of
temperature due to thermal diffusion. The Lagrangian derivative is indeed absent from
the asymptotic heat Eq. (\ref{eq:tempa0}). This property is crucial for the investigation
of small P\'eclet number regimes since numerical simulations are no
longer limited by the huge separation between the dynamical and diffusive time scales.

Another simplifying property appears when the P\'eclet-number equations (\ref{eq:vela0}),
(\ref{eq:tempa0}), (\ref{eq:diva0}), are written in terms of 

\[
\psi = \frac{\theta}{ P_e}.
\]
\noindent
Using this rescaled temperature deviation, the small-P\'eclet-number equations
become:

\begin{equation} \label{eq:velf}
\frac{\partial {\bf u}}{\partial t} + {\bf u}\cdot \nabla {\bf u} =
- \nabla p + R \psi {\bf e}_z + \frac 1{R_e} \nabla ^2{\bf u},
\end{equation}

\begin{equation} \label{eq:tempf}
w = \nabla ^2 \psi,
\end{equation}

\begin{equation}  \label{eq:divf}
\nabla \cdot {\bf u}=0,
\end{equation}

\noindent
where 
\begin{equation}
R=R_i P_e =\frac{t_{\rm D} t_{\kappa}}{{t_{\rm B}}^2}.
\end{equation}
\noindent
Although these
equations have been derived as a first order approximation 
of the Boussinesq equations in the limit of small P\'eclet number (see Sect. 2), 
they can also be interpreted as a zero order approximation of 
the Boussinesq
equations in the limit of small P\'eclet number provided the non-dimensional
number $R = R_i P_e$ is assumed to remain finite. Starting
from the dimensional Boussinesq equations,
one only has to use $P_e \Delta T_*$ as a reference temperature 
instead of $\Delta T_*$ and to assume Taylor-like expansions
of the form (\ref{eq:devv}) and (\ref{eq:devt}).
Then, provided $R$ remains constant, the above equations 
arise at the zero order in $P_e$.
The derivation presented in Sect. 2 has been preferred 
because it does not require the assumption of
an infinite Richardson number $R_i$.

The main interest of the system (\ref{eq:velf}), (\ref{eq:tempf}), (\ref{eq:divf}),
is that it only depends on two non-dimensional numbers,
$R$ and $R_e$. This is an important simplification as compared to
the original Boussinesq equations which are governed by three 
non-dimensional numbers, $R_i$, $P_e$ and $R_e$.
This simplification corresponds to the fact that
the amplitude of the buoyancy force is no longer
determined by two distinct processes, namely
the vertical advection against the stable stratification which produces temperature
deviations and the thermal diffusion which smoothes them out.
It is now determined by a single physical process which combines 
the effects of both processes..
We shall show below that this process is purely dissipative 
and that this dissipation is anisotropic (not effective for horizontal motions)
and faster for large scale motions.

To do so, we write down the kinetic energy conservation. Multiplying the 
momentum Eq. (\ref{eq:velf}) by the velocity vector and integrating over 
the whole domain, we obtain:
\begin{equation}
\frac{d E_{\rm kin}}{dt} =  R
\int_V w \psi \;dV - \frac{1}{R_e}
\int_V \epsilon \;dV + 
\int_S {\bf F}_{\rm kin} \cdot {\bf dS} .
\end{equation}
\noindent
where, on the r.h.s of this equation,  the first term is the work done by the 
buoyancy force, the second term represents the viscous dissipation into heat 
and the third term is the kinetic energy flux on the surface bounding the 
domain. Using Eq. (\ref{eq:tempf}), the work done by the buoyancy force can 
be divided into two terms to give:
\begin{equation}
R \int_V w \psi \;dV = - R \int_V
\left(\nabla \psi \right)^{2} \;dV 
+ R \int_S \psi \nabla \psi \cdot {\bf dS}.
\end{equation}
\noindent
As temperature deviations vanish on the bounding plates, the second
term also vanishes so that the kinetic energy 
conservation reduces to:
\begin{equation}
\frac{d E_{\rm kin}}{dt} = - R \int_V 
\left(\nabla \psi \right)^{2} \;dV
- \frac{1}{R_e} \int_V \epsilon \;dV 
+ \int_S {\bf F}_{\rm kin} \cdot {\bf dS}.
\end{equation}
\noindent
This equation shows that the combined effect of the stable stratification and 
the thermal diffusivity is purely dissipative. This simple result
has to be compared with the case of the Boussinesq equations
where the integrated work of the buoyancy could be positive or negative. 
As described in detail by Winters et al. (1995), it is then necessary to distinguish
the amount of kinetic energy which is irreversibly lost from the amount
of kinetic energy which has been transformed into potential energy
but can still return back to kinetic energy.
Here, the situation is simpler since all the kinetic energy extracted by the
buoyancy work is irreversibly lost.

In order to specify the time scale of this dissipative process, we rewrite the 
above equations without the non-linear terms and for inviscid motions restricted to a 
vertical plane (${\bf e}_{x}, {\bf e}_{z}$). The pressure term can first be eliminated using 
the incompressibility condition (\ref{eq:divf}).  Then, the two momentum equations are 
combined to eliminate the horizontal velocity and the simplified heat equation allows to 
eliminate the rescaled temperature deviation. Finally, the 
evolution of the vertical velocity is governed by:
\begin{equation} \label{eq:dispf}
\frac{\partial \Delta \Delta w}{\partial t} = R \frac{\partial^2 w}{\partial
x^{2}}.
\end{equation}
\noindent
Considering isotropic motions of length scale $l$, 
the time scale of this process is $ 1/(R l^2)$,
that is $t_{\rm B}^2 / t_{\kappa}$ in dimensional units.
It appears that this dissipative process is faster at large
scales than at small scales, which is just the opposite of what is observed in 
usual dissipative processes like thermal diffusion or viscous dissipation.
Here, however, the thermal diffusion does not act directly on the dynamics; it
affects the temperature deviations which in turn 
modifies the buoyancy force amplitude.
We have already seen that, in the limit of small P\'eclet numbers, rapid thermal
exchanges lead instantaneously to 
a balance between vertical advection against the mean stratification and thermal
diffusion. This balance is described by equation  (\ref{eq:tempf})} and it is
straightforward to show that the resulting amplitude of the temperature
deviations is stronger if the vertical velocity varies over a large length scale.
The amplitude buoyancy force is therefore stronger for velocity
fields varying over large length scales
and this explains why the combined effect of the stable stratification
and the thermal diffusion is faster at larger length scale.
Note that, while classical
dissipative processes are characterized by a Laplacian operator, the operator
of the present dissipation is the inverse of a Laplacian. This can be seen by
expressing $\psi$ as the inverse Laplacian of the vertical velocity and by
reporting this expression in the momentum equation. 

Another interesting property is the anisotropy of this dissipative process.
If one considers a velocity
field of the form $w \propto {\rm exp}[i(k_x x + k_z z)]$, where, 
as before, $k_z$ and $k_x$ represent 
its vertical and the horizontal scales, 
the characteristic time deduced from Eq. (\ref{eq:dispf}) is:
\begin{equation}
\tau = \frac{\left(k_x^2 + k_z^2 \right)^{2}}{R k_x^2},
\end{equation}
\noindent
which unsurprisingly corresponds to the inverse the damping rate
found in Sect. 2. 
The use of polar coordinates in the Fourier 
space is more appropriate to study the anisotropy of the process. With
$r^2 = k_x^2 + k_z^2$ and $tan(\alpha) = k_z/k_x$,
the time scale becomes:
\begin{equation}
\tau = \frac{1}{R} \left(\frac{r}{{\rm cos}(\alpha)}\right)^{2} 
\end{equation}
\noindent
where $\alpha = \pi/2$ corresponds to
horizontal motions and $\alpha = 0$ corresponds 
to vertical motions.

We observe that the dissipation acts primarily on vertical motions while purely
horizontal motions are not affected. This is not surprising since the 
buoyancy force only applies on the vertical component of the velocity.
What is more interesting is that this anisotropy is stronger 
than in the context of the non-diffusive
Boussinesq equations. Indeed, for a given value of
the wave vector modulus, the time scale $\tau$ increases faster towards 
horizontal motions ($\alpha \rightarrow \pi/2$) 
than the corresponding time scale of the buoyancy force
in a non-diffusive atmosphere $1/ \sigma_B = 1/({\rm cos}(\alpha) \sqrt{R_i}) $.
Considering motions strongly affected by the buoyancy force, we thus expect 
that these motions would be more predominantly horizontal in an atmosphere dominated 
by thermal diffusion than in a non-diffusive atmosphere.

\section{Discussion}

In this paper, we derived a small-P\'eclet-number approximation,
we discussed its validity for three flow examples and we analyzed its basic 
properties. 
In particular, we showed that the practical and theoretical 
difficulties characterizing the regime of very 
large thermal diffusivities and which had been mentioned in the introduction
are considerably simplified in the
context of the small-P\'eclet-number approximation.

In what concerns applications for the dynamics of stellar radiative 
zones, it must be stressed that some type of motions can not be investigated using
this approximation. First, there are no gravity waves in the context of the 
approximation whereas these waves could play an important role in
the radiative zone dynamics (Schatzman 1996). Second, thermal convective
motions penetrating the radiative zone boundary have a high P\'eclet number
so that the small-P\'eclet-approximation
is not suitable to investigate the overshooting layer
at the boundary with the thermal convective zone.

By contrast, there are various evidences that some motions contributing to the radial
transport of chemical elements and angular momentum are 
characterized by very small P\'eclet numbers and
could therefore be studied in the context of the small-P\'eclet-number approximation.
The fact that the thermal structure is determined by the radiative heat flux 
only shows that the P\'eclet number characterizing eventual radial motions is 
necessarily smaller than unity. But other observational constraints, obtained 
by measuring the surface abundance of chemical elements, give much smaller 
P\'eclet numbers (see a recent review by Michaud \& Zahn 
\cite{mich}). These P\'eclet numbers are defined as 
the ratio between the diffusion coefficient necessary to 
recover the observed surface abundance and the thermal diffusivity. In the
absence of more 
sophisticated models, this diffusion coefficient is assumed to represent
a vertical turbulent transport. For the sun, a P\'eclet number as small as $2 
\times 10^{-4}$ is obtained. This value may however be 
underestimated because the turbulence is most probably anisotropic. This aspect
is taken into account in the model of Spiegel \& Zahn (\cite{spie_2}) 
which describes the tachocline (the abrupt change in angular velocity at the 
top the solar radiative zone). The P\'eclet number which characterizes the 
horizontal turbulent motions and which is compatible with observed thickness of
the tachocline remains much smaller than unity ($P_e \approx 10^{-2}$). 

These estimates suggest to use the small-P\'eclet-number approximation to investigate
the property of small scale turbulent motions in stellar radiative zones. 
A first possible investigation could be the homogeneous turbulence in 
presence of a uniform mean shear and mean temperature gradient. With 
geophysical applications in mind, this configuration has already been 
extensively studied for large P\'eclet numbers (see Schumann 1996, for a review). 
A comparison with the small-P\'eclet-number case should be very 
instructive. Another important topic concerns the anisotropy between
vertical and horizontal motions in an atmosphere dominated by thermal
diffusion. Our linear study suggests that this anisotropy can be stronger
than in a non-diffusive atmosphere. The ratio between 
vertical and horizontal turbulent viscosities could be affected 
and this can estimated through numerical simulations of the small-P\'eclet-number
equations.

Before concluding, it must be noted that the limit of large diffusivities has 
already been considered (Spiegel \cite{spiegel}, Thual \cite{thual}) in the
context of the 
Rayleigh-B\'enard convection. However, an important physical property of 
thermal convection is lost in this limit. The thermal stratification 
is indeed assumed unchanged and this is not compatible with the general 
observation that convective motions transform the initial unstable 
stratification into an adiabatic stratification. There is no such an inconsistency
for the case of stably stratified radiative zones we considered here.


\begin{thebibliography}{}
\bibitem [1998]{can} Canuto V.M., Christensen-Dalsgaard J., 1998, 
Ann. Rev. Fluid Mech.
30, 167 
\bibitem [1997]{som} Cioni S., Ciliberto S., Sommeria J., 1997, 
J. Fluid Mech. 335, 
111
\bibitem [1974]{dudis} Dudis J.J., 1974, J. Fluid Mech. 64, 65
\bibitem [1996]{gough} Gough D.O., Leibacher J.W., Scherrer P.H.,
Toomre J., 1996, 
Science 272, 1281
\bibitem [1998]{moi} Ligni\`eres F., Califano F., Mangeney A., 1998,
Geophys. 
Astrophys. Fluid Dyn. 88, 81
\bibitem [1999]{moibis} Ligni\`eres F., Califano F., Mangeney A., 1999, 
to appear in A\&A
\bibitem [1998]{mich} Michaud G., Zahn J.-P., 1998, Theoret. Comput. Fluid
Dynamics 11, 183
\bibitem [1998]{pin} Pinsonneault M., 1998, ARA\&A
35, 557
\bibitem [1992]{thual} Thual O., 1992, J. Fluid Mech. 240, 229
\bibitem [1996]{schat} Schatzman E., 1996, J. Fluid Mech. 322, 355
\bibitem [1996]{schu} Schumann U., 1996, Dynamics Atmos. Oceans 23, 81
\bibitem [1962]{spiegel} Spiegel E.A., 1962, J. Geophys. Res. 67, 3063
\bibitem [1992]{spie_2} Spiegel E.A., Zahn J.-P., 1992, A\&A 279, 
431
\bibitem [1995]{win} Winters K.B., Lombard P.N., Riley J.J., D'Asaro E.R.,
1995, J. Fluid Mech. 289, 115 
\bibitem [1974]{zan} Zahn J.-P., 1974, in:
Ledoux et al. (eds.), Stellar instability and evolution, p. 185

\end{thebibliography}
\end{document}